\documentstyle[epsfig, twocolumn]{esapub}

\begin{document}

\def\halpha{H$\alpha$}
\def\microns{$\mu$m}
\def\msun{M$_\odot$}
\def\lsun{L$_\odot$}

%% Do not remove the following six lines:
\setlength{\parindent}{0pt}
\setlength{\parskip}{ 10pt plus 1pt minus 1pt}
\setlength{\hoffset}{-1.5truecm}
\setlength{\textwidth}{ 17.1truecm }
\setlength{\columnsep}{1truecm }
\setlength{\columnseprule}{0pt}
\setlength{\headheight}{12pt}
\setlength{\headsep}{20pt}
\pagestyle{esapubheadings}

%% Title - should be in capitals:
\title{\bf EMISSION-LINE DIAGNOSTICS OF GALAXY EVOLUTION WITH NGST}

%% If the author list spans more than one line then the {\bf (bold
%% font)} command must be inserted for each line
\author{{\bf R.C.~Kennicutt}  \vspace{2mm} \\
Steward Observatory, The University of Arizona, Tucson, AZ 85721, USA \\
Tel: 520-621-4032; FAX: 520-621-1532; Email: rkennicutt@as.arizona.edu}

\maketitle

\begin{abstract}
  As a byproduct of its search for the first star forming galaxies,
  NGST will obtain high-quality spectra for thousands of galaxies in
  the redshift range $z = 0 - 6$ and beyond.  Most of the galaxies 
  will possess strong emission-line spectra, and these spectra   
  will provide a wealth of quantititative information on
  their star formation rates, reddening, metal abundances, internal
  kinematics, and nuclear activity.  This will make it possible to
  construct a comprehensive picture of the buildup of stars, metals,
  dust, and the nuclei of galaxies with cosmological lookback time.
  This paper reviews what can be learned from the spectra of 
  distant emission-line galaxies.  Simulations of NGST spectra 
  are used to explore the instrumental tradeoffs for these applications,
  and the respective roles of NGST and groundbased surveys.
  \vspace {5pt} \\

%% Do not remove the previous commands. Your abstract should 
%% end with \vspace {5pt} \\  

%% Please insert your keywords here.
  Key~words: NGST; galaxy evolution; spectroscopy.

\end{abstract}

\section{INTRODUCTION}

One of the most important themes to emerge from this conference has
been that NGST, besides fulfilling its core goal of identifying the
first stars and galaxies, has the extraordinary potential to trace
the {\it physical} evolution of galaxies over the redshift range
$z = 0 - 6$ and higher (see the other papers in this volume by 
Dressler, Ellis, Stiavelli, and White).
The deep surveys with NGST will produce spectra for thousands of
galaxies, and these spectra will contain quantitative information
on the star formation rates (SFRs), reddening and extinction, 
metal abundances, kinematics, and nuclear properties for most of
these objects.  This will provide a comprehensive inventory of 
the buildup of stars, metals, dust, mass, and the nuclei of 
galaxies over lookback times ranging from a billion years after
the Big Bang to the present.  This goldmine of information may well prove
to be among the chief scientific achievements of NGST.  

\begin{figure}[!hb]
  \begin{center}
    \leavevmode
  \centerline{\epsfig{file=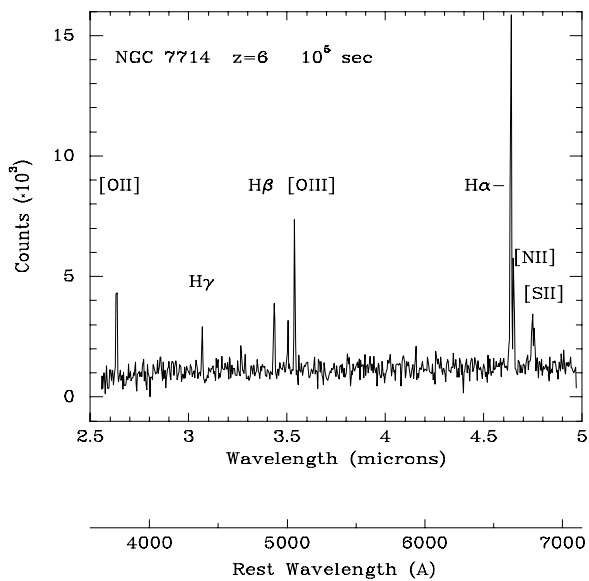,width=16cm}}
%  \vspace{6cm}
  \end{center}
  \caption{\em Simulated NGST spectrum of the nearby starburst galaxy
  NGC~7714, observed at redshift $z=6$.  The bottom axes show both
  observed (redshifted) and rest wavelengths, with the principal
  spectral features labeled.}  
  \label{fig:sample1}
\end{figure}

\begin{table*}[!ht]
  \begin{center}
    \caption{\em Summary of emission-line diagnostics.}
    \leavevmode

    \begin{tabular}[h]{llcc}
      \hline \\[-5pt]
      Parameter & Features & Resolution & S/N (\halpha) \\[+5pt]
      \hline \\[-5pt]
 SFRs  & H$\alpha$, ([OII], H$\beta$) & 100--300 & $>$10 \\
%   & [OII], H$\beta$ & & \\
   & & & \\
 Reddening/Extinction  & H$\alpha$, H$\beta$ & 500--2000 & $>$20 \\
   & & & \\
 Metal Abundance & H$\beta$, [OIII], [OII], (H$\alpha$, [NII], [SIII]) & 300--1000 & $>$10 \\
%   & H$\alpha$, [NII] & & \\
   & & & \\
 AGN Diagnostics & H$\beta$, [OIII], [NII], [SII], ([OI]) & 500--2000 & $>$20 \\ 
   & & & \\
 Kinematics & H$\alpha$, ([OII]) & 1000--10000 & $>$10--20 \\ 
% & [OII], [NII] & & \\
      \hline \\
      \end{tabular}
    \label{tab:table}
  \end{center}
\end{table*}

This lofty objective is achievable because NGST can provide 
high-S/N spectra of the complete visible and near-infrared spectral
regime, out to redshifts of at least $z=6$, thus making it possible
to apply the array of nebular diagnostic methods that have been
developed for groundbased studies of nearby galaxies.
As an illustration of NGST's unique potential,
Figure 1 shows a simulated NGST observation of the prototypical
nuclear starburst galaxy NGC~7714, observed at a redshift $z = 6$
(see Sec.~3 for a discussion of the simulation parameters).  
The combination of high spatial resolution, low background, and
continuous spectral coverage will make it possible to use spectra
like these to derive hard numbers on SFRs, dust content, metal
abundances, and nuclear emission properties for the  
emission-line galaxies that comprise the dominant population 
at $z > 1$.

The goal of this paper is to review the diagnostic methods
that can be applied to the integrated emission-line spectra of  
galaxies, and to explore the potential application of these
techniques with NGST.  I begin by discussing
the astrophysical parameters that can be extracted from emission-line
spectra, and the corresponding instrumental requirements in 
wavelength coverage, resolution, and S/N.  I then
present some examples of simulated NGST spectra, derived using
the baseline parameters in the NGST yardstick design in the
NGST ``black book" (Stockman 1997).  These provide useful insight
into the ranges of spectral resolutions and integration times that
are needed for these applications, and they clearly
illustrate the complementary roles that will be played by NGST
and future groundbased surveys.  My discussion is intended 
to complement the paper by Stiavelli presented elsewhere
in this volume.  In order to limit the scope of my presentation I
have emphasized applications of moderate-resolution ($R \le 2000$)
visible emission-line spectroscopy, and I will say relatively little
about the potentially important applications of absorption-line
spectra or near-infrared emission-line spectroscopy.

\section{EMISSION-LINE DIAGNOSTICS OF GALAXY EVOLUTION}

Figure 1 shows the principal nebular lines that are available
in the redshifted visible range.  The dominant features are
recombination lines of hydrogen (H$\alpha$, H$\beta$, H$\gamma$)
and helium (mainly He\thinspace I~5876), and forbidden
lines of [OII]$\lambda$3726,3729, [OIII]$\lambda$4959,5007,
[NII]$\lambda$6548,6583, and [SII]$\lambda$6717,6731.  
These lines by themselves provide a diverse array of diagnostics
of SFRs, reddening, abundances, velocity fields, and nuclear 
emission properties.  Because they span less than an octave in
wavelength the entire set of lines can be observed in the 1--5 \microns\
window over a wide redshift range ($z = 1.7 - 6.4$).  Much of the
diagnostic information is contained in lines with wavelengths of 
3727--5007 \AA, and this range is accessible shortward of 5 \microns\
out to $z \sim 9$.  Other useful visible-wavelength 
features include the shock and AGN-sensitive
[OI]$\lambda$6300,6363 doublet, and the temperature-sensitive 
auroral line of [OIII]$\lambda$4363.  However in most star forming
galaxies these features have fluxes of only a few percent of \halpha\ 
or less, and will only be seen in very high S/N spectra.  
The rest near-infrared region contains a number of other potentially
valuable diagnostic features, including the Paschen and Brackett
recombination lines (extremely valuable for measuring SFRs in
dusty regions), the ionization-sensitive [SIII]$\lambda$9069,9532
doublet, the shock-sensitive [FeII] lines longward of 1.2 \microns,
and several diagnostics of molecular and coronal-phase gas.
These features will be redshifted into the zodiacal background 
for high-redshift objects, but will provide valuable diagnostic
information for intermediate-redshift galaxies.

Table 1 presents a summary of the observational requirements for 
some of the main nebular diagnostics.  In each 
case I list the relevant emission lines, and the approximate ranges of 
spectral resolution and S/N required, with   
higher values indicating the optimal desirable
performance, and the lower values indicating what I
regard as the minimal capability needed for quantitative applications.
I briefly discuss each application individually below.
Since observations of galaxy kinematics are dealt with extensively
in other papers, I will not discuss that application further here.

\subsection{Star Formation Rates}  

The \halpha\ luminosity of a galaxy provides
a robust measure of its global SFR, which is directly tied to the luminosity
and mass of the young massive stellar population (Kennicutt 1983, 1998).
\halpha\ data provide most of our information on the systematics of
SFRs in nearby galaxies, but the redshifting of \halpha\ outside the
visible window for $z \ge 0.5$ has hampered its application 
to high-redshift galaxies.  Instead, most of what is known about
the cosmic evolution of the SFR comes from redshifted UV continuum
measurements (e.g., Madau et al.~1996), supplemented to some extent by 
[OII] emission-line observations (e.g., Cowie et al.~1997).  As discussed
at this conference by Madau and others, large uncertainties in the
extinction corrections in the UV continuum measurements introduce
uncertainty into both the normalization and shape of the Madau diagram.
NGST will be capable of detecting \halpha\ directly out to $z \sim 7$,
with reddening information available from the Balmer decrement (below).
This will provide independent measures of the SFR, and a solid cross-check
on the cosmic evolution of the SFR determined from other methods.  

\begin{figure*}[!ht]
  \begin{center}
    \leavevmode
  \centerline{\epsfig{file=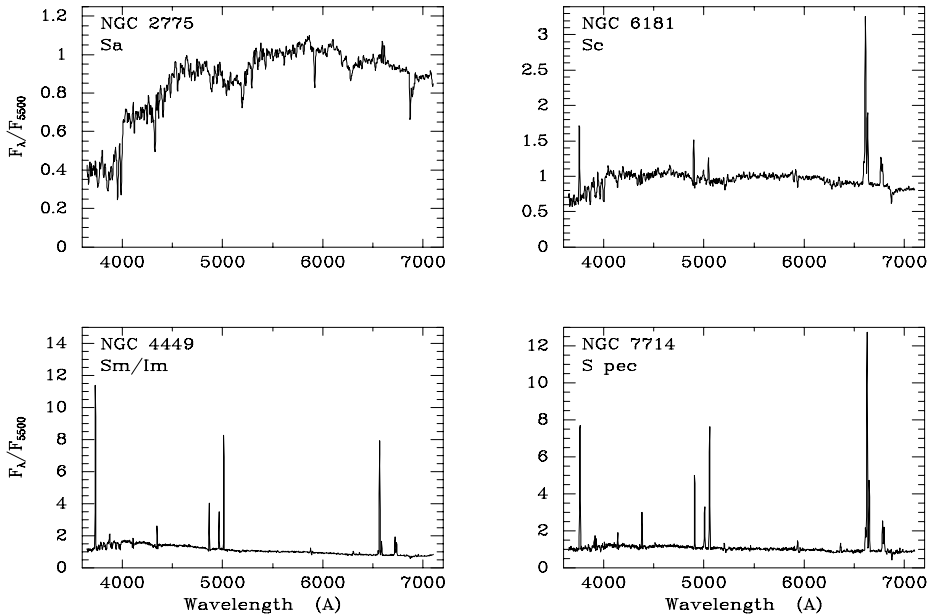,width=16cm}}
% \vspace{4.5cm}
  \end{center}
  \caption{\em Integrated spectra of 4 nearby spiral and irregular 
    galaxies, from the sample of Kennicutt (1992b). }
  \label{fig:sample2}
\end{figure*}

The \halpha\ line is usually among the strongest emission features in
the integrated spectra of galaxies, so this application can 
be accommodated with relatively modest spectral resolution and S/N.
Figure 2 shows examples of moderate-resolution ($R = 800$) integrated 
spectra for 4 nearby galaxies from the atlas of Kennicutt (1992b), 
chosen to span the range of SFRs observed in local spiral and irregular
galaxies.  A minimum resolution is set by the need to resolve \halpha\
from the [NII]$\lambda$6583 line ($R \ge 400$), but useful information 
on the SFR
can be obtained even with resolutions of order 100 and S/N of $\sim$10.

Exploratory studies of the redshifted \halpha\ emission of distant galaxies are
already being carried out from the ground (e.g., Bechtold et al.~1997),
and several large surveys are being planned for groundbased spectrographs
on 8--10 m class telescopes.  The main limitation in these studies is
not sensitivity but rather interference from the telluric OH emission
in the 0.8--2 \microns\ region, and the steeply rising thermal background 
longward of 2 \microns.  It is likely that \halpha-determined 
SFRs will have been measured for thousands of galaxies out to $z \sim 2$
in the next decade, as well as for a smaller number of luminous objects at
higher redshifts.  The main value of NGST will be in extending
the application to higher redshifts, and in obtaining SFRs for complete
samples over the full redshift range.

In applications where \halpha\ is unobservable it is possible to
use the higher-order Balmer lines (e.g., H$\beta$) or even a forbidden
line such as [OII]$\lambda$3727 as a substitute SFR measure.  
The H$\beta$ line is useful in galaxies with sufficiently
strong emission so that the effects of underlying stellar absorption
are small.  This condition is satisfied in local emission-line starburst
galaxies, but not among normal disk galaxies or
post-burst galaxies (Kennicutt 1992a).  
Several workers have used the [OII] line
to estimate the evolution in the SFR with redshift (e.g., Cowie
et al.~1997).  The advantages of [OII] are its relative strength 
and its observability to much higher redshifts than \halpha.
However the strength of the line is not directly tied to the
photoionization rate; the ratio of the line flux to the
SFR is sensitive to excitation variations as well as to reddening.
This ratio can easily vary by a full order of magnitude in luminous
spiral galaxies (Kennicutt 1992a), but it appears to be a more
reliable measure of the SFR in strong emission-line galaxies
(Gallagher et al.~1989, Jansen et al., private communication).
Several large spectrophotometric surveys of nearby galaxies
that are currently under way should provide a firmer test
of the reliability of [OII]-derived SFRs over the next few years.
However the systematic changes in [OII] excitation with abundance,
dust content, and ionization parameter remain a cause of some
concern for lookback applications (Kennicutt 1992a, 1998), and
whenever possible a SFR scale directly anchored on recombination
lines is preferable.  

\subsection{Reddening and Extinction}

As mentioned earlier, extinction remains as the largest
single source of uncertainty in current studies of the evolution of
the cosmic star formation rate.  Nebular SFRs are hardly
immune from extinction problems, but the nebular spectrum offers a 
means of constraining the reddening and extinction via the Balmer
decrement.  This requires considerably higher S/N spectra than for
\halpha-based SFRs alone, ideally with sufficient spectral resolution
and S/N to detect any stellar absorption under H$\beta$.  

Spectra integrated over a large spatial region will tend to underestimate
the total extinction, because the observed Balmer decrement will
be weighted to the regions of lowest line-of-sight extinction.
However studies of nearby starbursts and disk HII regions suggest that the
reddening measured by the Balmer decrement can provide a reasonable
estimate of the visible and UV extinction (e.g., Calzetti et al.~1994,
Buat \& Xu 1996).

One regime in which the visible measurements break down entirely is
in the dusty cores of luminous infrared starburst galaxies.  Most of
these objects show a pronounced emission-line spectrum in the visible,
but the nebular reddening may bear little relation to the 
extinction, which often amounts to several magnitudes in the visible.
Such objects may represent a substantial component of the high-redshift
starburst population, and thus it is important to supplement any
survey of the redshifted UV--visible spectra with coordinated observations
of the redshifted infrared emission.  In cases where the Paschen lines
are detectable with NGST they would provide a critical test of the 
reliability of the \halpha\ and UV-derived SFRs.

\subsection{Metal Abundances}

One of the most exciting prospects for NGST spectroscopy is the
determination of gas-phase metal abundances for high-$z$ galaxies.
As discussed in the review by Stiavelli, a rigorous determination
of the nebular oxygen abundance requires an accurate measurement
of the temperature-sensitive [OIII]$\lambda$4363 line or another
auroral line.  However the flux of this line is typically of
order 0.1--1\%\ of [OIII]$\lambda$4959,5007, so direct
abundance determinations will be limited to the brightest 
galaxies, and those with relatively weak stellar continuum.

Most determinations of abundances in high-$z$ galaxies will probably
be based on the empirically-calibrated strong-line indices, for
example the widely applied $R_{23}$ index of Edmunds \& Pagel (1984):

\begin{equation}
 {{ R_{23}}} \equiv { {[OII]\lambda3726,3729 + [OIII]\lambda4959,5007} \over {H\beta} }
\label{eq:equation1}
\end{equation}

For metal abundances $Z \ge 0.2 Z_\odot$ the relationship between 
$R_{23}$ and oxygen abundance is monotonic, with the forbidden
lines becoming {\it weaker} with increasing abundance.  The
index has been calibrated using a combination of observed 
abundances (calibrated with [OIII]$\lambda$4363 measurements)
and nebular models for the most metal-rich regions (e.g., McGaugh 1991).  
Over the
abundance range of about $0.2~Z_\odot < Z < Z_\odot$ the $R_{23}$
index can provide HII region abundances that are accurate to 
within about $\pm$0.1--0.2 dex, with somewhat larger errors at
higher abundances, where the calibration rests almost entirely on models.  
The method has been widely applied to HII region
surveys of nearby galaxies (e.g., Zaritsky et al.~1994), and in
principle should be applicable to high-$z$ objects.  Kobulnicky \& Zaritsky
(1998) have used Keck spectroscopy of a sample of galaxies at $z = 0.1 - 0.5$
to test this technique, and to explore the physical nature of the 
compact narrow emission-line galaxies that are frequently observed
at these redshifts (e.g., Koo et al.~1995).  The abundances measured
for these objects, typically in the range 0.2--1.5 $Z_\odot$, already
rule out a popular interpretation of these objects as progenitors to
present-day dwarf spheroidal galaxies.  This illustrates the potential
power of abundance measurements for constraining the physical evolution
of high-redshift objects, especially if the data can be combined with
measurements of their luminosities, 
velocity dispersions, SFRs, and other properties.

\begin{figure}[!ht]
\begin{center}
\leavevmode
\centerline{\epsfig{file=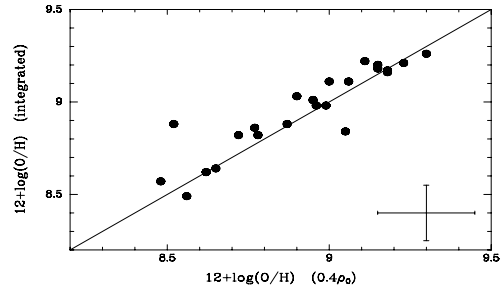,width=8.0cm}}
\end{center}
\caption{\em Comparison of oxygen abundances inferred from the integrated
emission-line spectra of disks, using the $R_{23}$ method, and the 
actual abundance at 0.4 $R_0$, from the study of Kobulnicky et al.~(1998).}
\label{fig:sample3}
\end{figure}

A major concern for the application of these methods to distant galaxies
is the effect of beam smearing on the integrated line ratios.  The 
strengths of the [OII] and [OIII] lines follow well-defined but 
nonlinear dependences on abundance, and it is not at all clear
{\it a priori} whether the nebular excitation inferred from a large spatial
region--- with contributions from several regions with possibly 
different abundances and ionization properties--- will necessarily 
correlate with the mean abundance in the same way as for individual
HII regions.  I have investigated this recently in collaboration with
Chip Kobulnicky and James Pizagno (Kobulnicky et al.~1998).
We combined published HII region spectra of a large sample of galaxies with
well-measured abundance distributions with radial 
\halpha\ emission profiles, to simulate the integrated emission-line
spectra of the galaxies.  We then compared the abundance one would
infer from the integrated $R_{23}$ ratio with the actual abundance 
measured for the disk.   Figure 3 shows the comparison, with the
integrated measurement compared in this case to the observed abundance
at 0.4 $R_0$.  There is an excellent correlation, with only one
pronounced outlier (M101), a galaxy with an unusually
extended star forming disk and an unusually strong abundance gradient.
This comparison suggests that the sampling error in determining 
mean abundances from the integrated spectra is comparable to or smaller
than the intrinsic error in the $R_{23}$ method itself, so comparisons
like those carried out by Kobulnicky \& Zaritsky (1998) should be valid.
The comparison shown in Figure 3 does not include the contribution of 
diffuse ionized gas to the integrated spectrum, but observations by 
Hunter (1994) and Martin (1997) suggest that while the relative strengths
of [OII] and [OIII] can be quite different from those of normal HII regions,
the $R_{23}$ values roughly follow the same abundance relation to first
order.  We have also investigated the effects of beam smearing on
spectra with detectable [OIII]$\lambda$4363, using measured longslit
spectra of nearby galaxies.  There is evidence for a systematic bias
toward lower inferred abundances, but the effect is small ($\sim 0.1$ dex),
and consistent enough that a correction for this effect can be applied.

This technique is already being applied from the ground, for low-redshift
galaxies in the visible (e.g., Kobulnicky \& Zaritsky 1998) and for
$z \simeq 3$ galaxies in the infrared (e.g., Pettini et al.~1998).
It is likely that such measurements will be carried out for large
samples of galaxies before NGST is launched.  However the groundbased
observations will be handicapped in several important respects.
At a minimum, a reliable abundance measurement requires the measurement
of at least three emission features ([OII]$\lambda$3727, H$\beta$, 
and [OIII]$\lambda$4959 or [OIII]$\lambda$5007), and the probability of
avoiding a strong telluric emission or absorption line quickly
decreases as the number of lines increases.  Equally serious
is the breakdown in the $R_{23}$ index below abundances of roughly 
0.2 $Z_\odot$.  At very low abundances the forbidden-line strengths
decrease with decreasing abundance, so $R_{23}$ becomes a 
double-valued function of oxygen abundance, and there is a broad
region between about 0.1--0.3 $Z_\odot$ where $R_{23}$ is roughly
constant, and thus is completely insensitive to abundance.
Unfortunately this is just in the range that one might expect many
high-redshift galaxies to fall.  One can circumvent this problem
either by observing a temperature-sensitive line such as [OIII]$\lambda$4363,
or more realistically by combining the information on $R_{23}$ with
the ratio of [OII]/[OIII] and measurements of other lines such as
[NII]$\lambda$6583, [SII]$\lambda$6717,6731, and/or [SIII]$\lambda$9069,9532
(e.g., Skillman 1989).  Obtaining spectra over such a wide spectral
range will be problematic from the ground, but is feasible with NGST
(see Figure 1).  

The instrumental requirements for abundance measurements are similar
to those for measuring SFRs and reddening, except that a broad simultaneous
spectral coverage is desirable in this instance.  A clean resolution
of [NII] from \halpha\ is desirable, especially in low-metallicity objects
where [NII] will be very weak.  This implies spectral resolution 
$R \ge 500$ and S/N $>$ 30 in the strongest lines.  For more metal-rich
objects the requirements are somewhat lower.

% \subsection{Internal Kinematics}
% 
% Several other contributors to this conference have discussed applications
% of velocity measurements of distant galaxies, so I will only briefly
% mention this important application for completeness.  The high spatial
% resolution of NGST will produce rotation curves for large numbers of
% galaxies, for moderately high spectral resolution ($R \ge 2000$).
% Useful information may be extractable from lower-resolution spectra
% if the S/N is high enough to permit reliable line centroiding.
% Measuring the kinematics and velocity dispersions of low-mass 
% galaxies requires much higher resolution, ideally up to $R = 10000$.
% Whether such performance is feasible for NGST remains to be determined.
% 

\subsection{Nuclear Activity and Star Formation}

Another area in which NGST should make major progress is in tracing the
evolution of nuclear activity and star formation with lookback time.
For most realistic cosmologies the angular resolution of NGST 
corresponds to linear scales of a few hundred parsecs (for all redshifts
of interest), and galactic nuclei will be cleanly resolved in images
and spectra.  The same diagnostics as discussed in Sec.~2.1 can be
used to measure the nuclear SFRs, and compare the cosmic evolution
of the nuclear and global SFRs.  
The extensive wavelength coverage that is attainable with NGST
will also make it possible to apply the familiar diagnostic lines
([OIII], [NII], [SII], [OI], [SIII]) to distinguish
star forming nuclei from Seyfert nuclei, LINERs, and other AGN types
(e.g., Veilleux \& Osterbrock 1987).  

This application has not been widely discussed in framing the science case
for NGST, but I believe it represents an enormously rich opportunity.
Spectroscopic observations of nearby galaxies suggest that massive nuclear 
black holes may be a ubiquitous phenomenon, at least among galaxies with 
large central spheroidsl.  If this is the case, NGST offers us the 
opportunity to directly observe the buildup of these central mass
concentrations, and extend the study of the frequency and luminosity
evolution of nonthermal nuclear activity to a millionfold dynamic
range in nuclear luminosity.  All of these applications 
will be nearly impossible to duplicate from the ground.     

The instrumental requirements closely match
those for the abundance measurements, in cases where only the fluxes
of emission lines are required.  At resolutions $R \ge 1000$ and
high S/N one can begin to use the nuclear line profiles as 
diagnostics as well.

\section{PROSPECTS WITH NGST}

The previous section outlined the astrophysical
and instrumental wish lists for these applications.  What are the
prospects for meeting these with NGST?  To address this question I 
have constructed simulations of spectroscopic observations with NGST,
using integrated spectra of nearby galaxies from Kennicutt (1992b)
and the artificial data package in IRAF.  Slit spectra were constructed
for selected galaxies observed at redshifts $z = 0 - 9$.  I also  
used \halpha\ images of a few nearby galaxies to simulate
2D emission-line images in the NGST focal plane, in order to explore
the effects of spatial sampling on the observability of distant galaxy
spectra.

\begin{table}[h]
  \begin{center}
  \caption{\em Simulation parameters.}
  \leavevmode

  \begin{tabular}[h]{ll}
    \hline \\[-5pt]
 Cosmology  & H$_0$=75, $\Omega_0$=0.3, $\Lambda_0$=0 \\
 Galaxy Luminosity  & $M{_B^*}$ = $-$21.0 \\
 NGST Aperture & $D$ = 8 m\ \  ($A$ = 50 m$^2$) \\
 System Efficiency & 0.2 \\
 Sky Background & 2$\times$ DIRBE minimum \\
 Instrumentation & \\
 \quad -- pixel size & 0.03 arcsec \\
 \quad -- readout noise & 10 $e^-$ \\
 \quad -- dark noise & 0.01 $e^-$/sec/pixel \\
 \quad -- integration time & 1000 sec \\
 Spectroscopy & \\
 \quad -- slit width & 0.1 arcsec  \\
 \quad -- slit length & 0.1--0.3 arcsec ($\sim$1--3 kpc) \\
 \quad -- resolution & $R$ = 800  (3 pixels) \\
 \hline \\
 \end{tabular}
 \label{tab:table}
 \end{center}
\end{table}

Table 2 lists the parameters and assumptions that were used in the
simulations shown here.  It is important to bear in mind that with
so many uncertainties in the instrument design one can only approximate
what might be expected with NGST, and in most respects I have attempted
to be conservative in my assumptions.  As discussed later, the
most critical parameters are the detector noise figures, which in this
case were taken from Stockman (1997).  Almost all 
spectroscopic applications are detector noise limited, and this
influences many of the observational parameters, in particular
the optimal slit sampling and the integration times.  For the parameters 
chosen, detector read noise and dark noise are significant contributors.
In the 1--5 \microns\ region the assumed sky background 
is typically less than 10\%\ of the dectector noise, and can be 
ignored for most applications (however the sky quickly dominates the
noise longward of 5 \microns).  To be conservative I have
subtracted a background from each spectrum, which has the practical
effect of increasing the net noise by a factor of 1.414.  

\begin{figure*}[!ht]
  \begin{center}
    \leavevmode
  \centerline{\epsfig{file=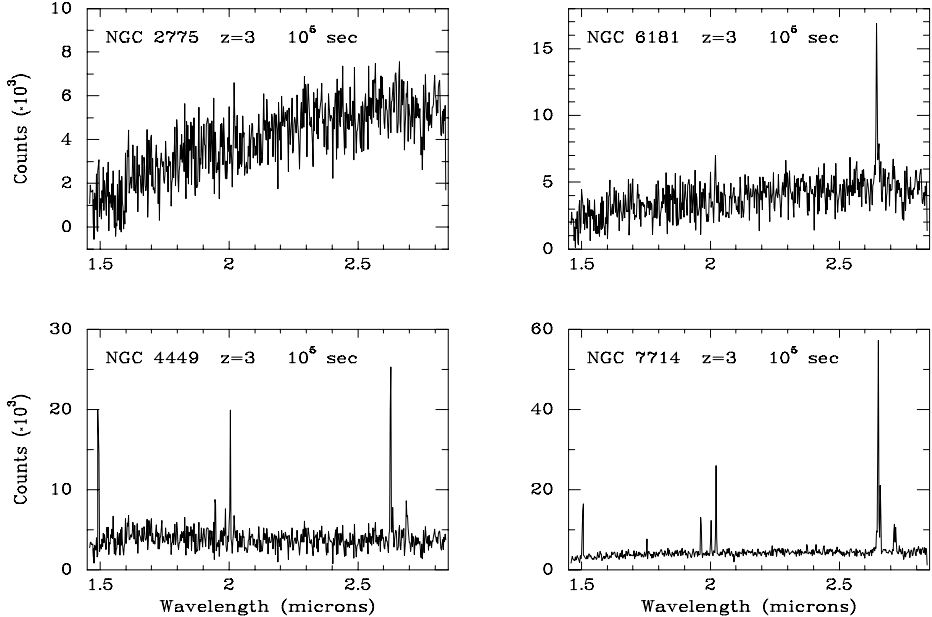,width=16cm}}
% \vspace{4.5cm}
  \end{center}
  \caption{\em The same galaxies as in Figure 2, but observed at 
  $z = 3$ with NGST, using the paramters in Table 2 and an assumed
  integration time of 100 ksec.  The vertical scale is thousands of
  counts per channel.}
  \label{fig:sample4}
\end{figure*}

As an illustration of the range of spectra that are likely to be 
seen with NGST, Figure 4 shows the same nearby galaxies
as in Figure 2, but in this case observed at $z = 3$ with NGST,
with spectral resolution $R = 800$ and a total integration time
of 100 ksec.  To facilitate the comparison the galaxy luminosities
have been normalized to a common value of $M_B = -21$, but it 
is assumed that only 20\%\ of this luminosity is contained in
an aperture of 0.1$\times$0.3 arcsec.  Larger apertures would 
include more of the disks, but would degrade the S/N due to the
larger number of detector pixels. 
For NGC~7714 most of the emission is contained in the inner 1 kpc,
so I have used a 0.1 arcsec square aperture for this object.

Figure 4 shows that the most useful emission-line spectra will be
obtained for galaxies that are analogs to the most active nearby
star forming galaxies, such as NGC~4449 and the starburst galaxy
NGC~7714.  One can expect to detect \halpha\ in the analogs to
present-day Sb--Sc spirals, but the other diagnostic lines will 
not be cleanly detected.  In the most evolved systems even \halpha\
may be absent.  However this limitation is less serious than would
appear from Figure 4, because galaxies with stellar populations as
evolved as those in NGC~2775 and NGC~6181 will be rare above redshifts
$z \sim 2$; indeed we already know from groundbased lookback studies
that the average emission-line strength in galaxy spectra increases
sharply already from $z = 0$ to 1 (e.g., Cowie et al.~1997).
If current observations are an accurate indication, emission-line
galaxies like NGC~4449 and NGC~7714 will be the dominant population
at high redshift, and many of these objects will have luminosities
which are more than an order of magnitude higher than the examples
shown here.  

\begin{figure}[!hb]
  \begin{center}
    \leavevmode
  \centerline{\epsfig{file=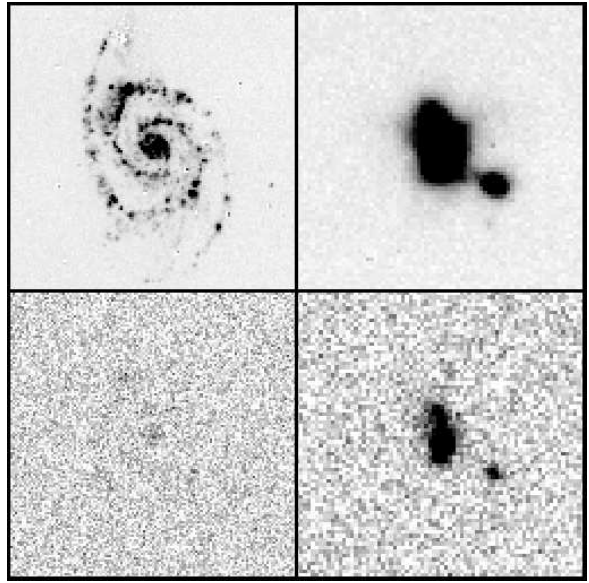,width=8cm}}
% \vspace{4.5cm}
  \end{center}
  \caption{\em Images showing the distribution of \halpha\ emission
  in M51 (left) and the starburst merger NGC~1222 (right).  The
  upper panels show the galaxies as they are seen locally, but
  degraded to the spatial resolution of NGST at redshift $z = 2$.
  The lower panels show the actual detected emission distributions
  at $z = 2$, for 100 ksec stacked exposures and the instrument
  parameters given in Table 2.  The panel size is roughly 3.6 arcsec
  (40 kpc) for M51 and 1.8 arcsec (20 kpc) for NGC~1222.}
  \label{fig:sample5}
\end{figure}

\begin{figure*}[!ht]
  \begin{center}
    \leavevmode
  \centerline{\epsfig{file=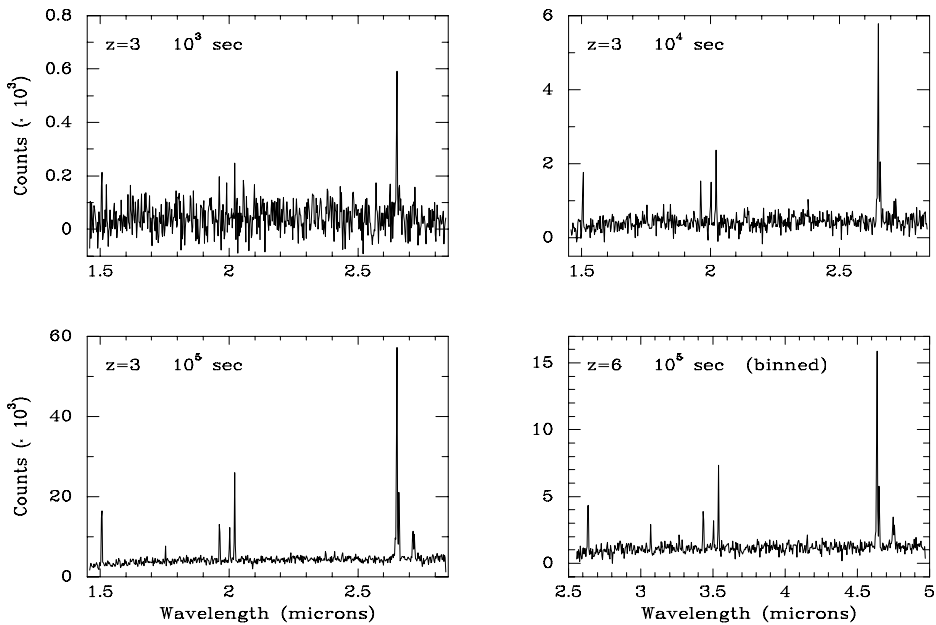,width=16cm}}
% \vspace{4.5cm}
  \end{center}
  \caption{\em Comparison of spectra of NGC~7714 at $z = 3$, for different
  total integration times.  The bottom right panel shows a spectrum at
  $z = 6$, with optimized on-chip binning to reduce effective readout noise.}
  \label{fig:sample6}
\end{figure*}

Another factor that strongly influences the observability of emission-line
spectra with NGST is the spatial distribution of the star formation.
This is illustrated in Figure 5, which shows simulations of \halpha\
monochromatic images of M51, a relatively high surface brightness star
forming spiral, and NGC~1222, a merger remnant with a very luminous
circumnuclear starburst.  
The upper panels show \halpha\ images of
each galaxy, degraded to the 0.03 arcsec pixel sampling of the NGST
baseline design (for redshift $z = 2$, but the angular scale is
nearly independent of redshift for $z > 1$).  The panel sizes are 
approximately 120 pixels (3.6 arcsec) square for M51, and 60 pixels
(1.8 arcsec) for NGC~1222;  
this illustrates the superb spatial resolution that
will be offered by NGST.  The lower panels 
show simulated 100 ksec stacked exposures with NGST for $z = 2$.
Although the total SFRs of the two galaxies are comparable, the
emission lines in M51 are barely detectable above $z = 2$, because
the star formation is spread over several thousand detector pixels.
The signal can be recovered to some extent of course by binning
the data, but this eliminates much of the advantage of NGST over
groundbased applications; indeed a comparable simulation of M51
using a groundbased 8 m telescope (not shown) shows that for
resolutions of order a few tenths of an arcsecond the groundbased
observations are competitive with NGST shortward of 2 \microns.  This is 
an important lesson, namely that for observations of analogs of 
local spirals at modest redshifts, much of our information is likely
to come from groundbased surveys, and the main contribution of NGST
will be to study high surface brightness regions in these galaxies
(e.g., their nuclei) at higher spatial resolution.

For compact star forming galaxies (e.g., NGC~1222, NGC~7714), the
spatial resolution of NGST is used to full advantage, and simulations
show that the bright emission lines can easily be detected in 100 ksec
exposures out to $z \ge 9$.  Recent HST observations suggest
that most of the star formation seen at high redshift in fact takes
place in very compact objects, and if this interpretation is correct
(i.e., not a consequence of surface brightness bias), then NGC~1222
is the most relevant analog to high-$z$ galaxies, and NGST will 
easily obtain high S/N spectra of these objects to impressively
high redshifts.  It is worth noting that most local examples of 
these luminous starburst galaxies are characterized by very dense
dusty star forming regions, so extinction corrections are essential
for interpreting the properties of these objects.

The simulations are also useful for exploring the effects of
different instrumental tradeoffs on the quality of the spectra.
For example Figure 6 compares NGST spectra of NGC~7714 at $z = 3$,
for integration times of 1, 10, and 100 ksec.  Even for this
unusually luminous and compact emission-line galaxy, useful 
spectral diagnostics require a minimum exposure time of $\sim$3 hours,
so for applications of this type, integrations of order a day or
longer will be needed to make a significant advance over 
contemporary groundbased
surveys.  For the parameters used here the
exposure times required are tightly linked to the detector noise properties,
and reduction in the detector noise could dramatically lower the
exposure times required.  This is illustrated in the bottom right
panel of Figure 6, which shows a 100 ksec exposure of NGC~7714
at $z = 6$, but with an optimal use of on-chip binning to reduce
the readout noise.

Figure 7 shows a similar comparison, but in this case for different
spectral resolutions ranging from $R = 100$ to 2500.  These are based on 
identical detector parameters, so they do not include the incremental
gain in S/N that can be obtained with on-chip binning at low resolution.
However this approximation
does not affect the interpretation of the results.  Useful information
on the strong lines can be obtained down to a limiting resolution of
$R = 300 - 400$, but even at this resolution critical diagnostics
such as [NII] are lost.  Resolutions in the range $R = 800 - 1200$
appear to provide the best balance between resolution and S/N,
at least for this particular application.

\begin{figure*}[!ht]
  \begin{center}
    \leavevmode
  \centerline{\epsfig{file=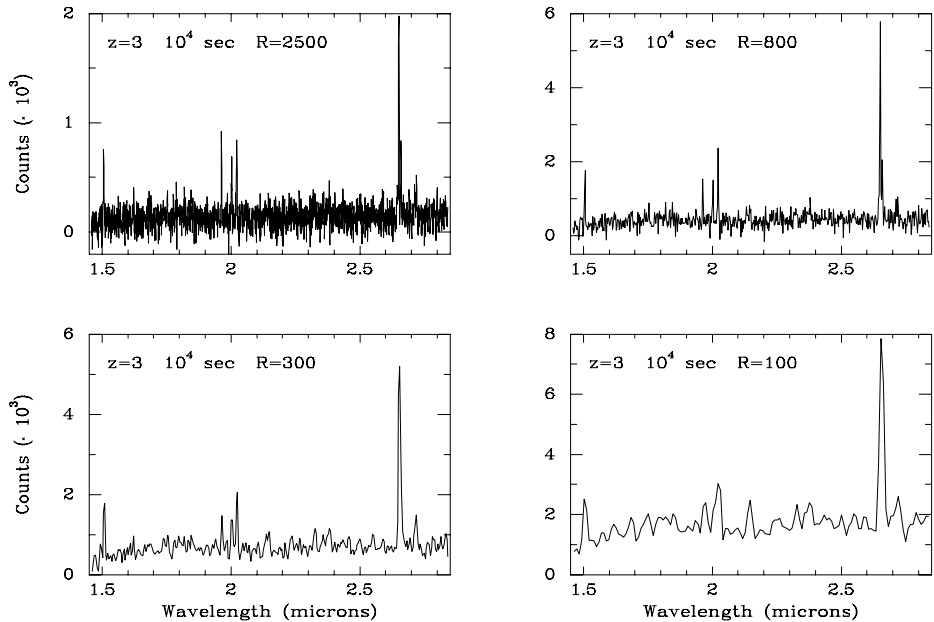,width=16cm}}
% \vspace{4.5cm}
  \end{center}
  \caption{\em Comparison of spectra of NGC~7714 at $z = 3$, for
  different spectral resolutions.}
  \label{fig:sample7}
\end{figure*}

\section{LESSONS AND CONCLUSIONS}

I hope that this paper has made a convincing case for the enormous
potential that NGST has for reconstructing a comprehensive astrophysical
picture of galaxy formation and evolution.  There is little doubt that
the armada of surveys with 8--10 m telescopes to be carried out over the next
decade will provide the basic foundation of this picture, but
it remains for NGST to provide the complete set of 
hard data on galaxy masses, star
formation rates, gas and dust contents, chemical abundances, and
nuclear properties that will result in a full physical understanding
of the origins of galaxies, stars, galactic nuclei, and the chemical
elements.  

Almost all of this science can be carried out within the NGST baseline 
capabilities (4--8 m telescope, 1--5 \microns\ wavelength coverage).
Exploiting this opportunity, however, imposes instrumental drivers 
that may be somewhat more stringent than required by the core mission
of redshift surveys of the highest-redshift galaxies.  In particular,
obtaining diagnostic-quality spectra drives the spectroscopic 
instrumentation towards resolutions of order $10^3$ and signal/noise
of order tens.  This probably entails typical observing times of order
a day (or longer) rather than hours, although this requirement is
a direct function of detector noise properties, and improvements in
this area would have enormous benefit for all high-$z$ spectroscopic
applications.  If these needs can be met, the astrophysical payoff
promises to be quite extraordinary.

\section*{ACKNOWLEDGMENTS}

It is a pleasure to acknowledge U of A graduate students
Audra Baleisis and Anne Turner for furnishing the \halpha\ image 
of NGC~1222, and undergraduate student James Pizagno for his help
with the abundance calculations presented here.  I also thank
Richard Ellis, Marijn Franx, Garth Illingworth, Chip Kobulnicky, 
Hans-Walter Rix,
Claudia Rola, Matthias Steinmetz, Peter Stockman, and Rodger Thompson
for useful discussions about this subject.
Some of the work described here is supported by the U.S. National
Science Foundation through grant AST-9421145.

\end{document}